# Correlation between Organic Magnetoresistance (OMAR) and Ferromagnetic ordering

Sayani Majumdar[1,2*], Himadri Majumdar[1], Jan-Olof Lill[3], Johan Rajander[3], Reino Laiho[2] and Ronald Österbacka[1]

[1]*Centre for Functional Materials and Department of Physics, Åbo Akademi University, 20500 Finland.*

[2]*Wihuri Physical Laboratory, Department of Physics and Astronomy, University of Turku, 20014 Finland.*

[3] *Accelerator Laboratory, Turku PET Center, Åbo Akademi University, 20500 Finland.*

[*]**Corresponding author: sayani.majumdar@utu.fi**

## Abstract

We report observation of ferromagnetic (FM) ordering in organic semiconductors, namely regio-regular poly (3-hexyl thiophene) (RRP3HT) and 1-(3-methoxycarbonyl)propyl-1-phenyl-[6,6]-methanofullerene (PCBM), in the temperature range of 5 – 300 K in addition to magnetoresistance (OMAR) observed in the diodes made from the same materials. Particle induced x-ray emission spectroscopy confirms the presence of dilute magnetic impurities in the materials mainly as residues from the synthesis process. However, upon blending these two materials with FM signal, the FM ordering is suppressed by a huge paramagnetic (PM) signal indicating ground state charge transfer formation in the blend. Together with the magneto-transport studies, these results indicate that OMAR response is observed in a device only when the corresponding active materials are FM. In the diodes with P3HT:PCBM complex, that as a blend shows PM response, OMAR vanishes almost completely. We propose that ferromagnetism in the active material can have important correlation with the OMAR response in the diodes.

**Intrduction**

Over the past two decades, the growth of organic electronics has been phenomenal [1-3]. However, since recently, all research has been predominantly focused towards the understanding of charge transport in organic semiconductors (OS) small molecules and pi-conjugated polymers (PCP)s and its implications in devices. Study of spin in organics is a relatively unexplored territory. It provides a whole new scope of physics along with immense potential for application. Spintronics (spin-based electronics) refers to the study of the role played by electron spin in solid state physics, and possible devices that specifically exploit spin properties instead of or in addition to charge degrees of freedom [4].

Major challenges in the field of spintronics are addressed by experiment and theory pertaining to the optimization of electron spin lifetimes, the detection of spin coherence in nanoscale structures, transport of spin-polarized carriers across relevant length scales and heterointerfaces, and the manipulation of both electron and nuclear spins on sufficiently fast time scales. It is envisioned that the merging of electronics, photonics, and magnetics will ultimately lead to new spin-based multifunctional devices such as spin-FET (field effect transistor), spin-LED (light-emitting diode), spin RTD (resonant tunneling device), optical switches operating at terahertz frequency, modulators, encoders, decoders, and quantum bits for quantum computation and communication. The success of these ventures depends on a deeper understanding of fundamental spin interactions in solid state materials as well as the roles of dimensionality, defects, and semiconductor band structure in modifying these dynamics. With proper understanding and control of the spin degree of freedom in semiconductors and ferromagnet-semiconducting heterostructures, potential for high-performance spintronic devices is excellent.

In recent years semiconductors are studied widely as promising spin transport materials [5] for spintronic applications. Although great research effort is dedicated to explore the inorganic magnetic semiconductors for spintronic applications very little attention has been paid to the use OS molecules and PCPs as spintronic materials. Long spin-correlation length is expected in OS compared to their inorganic counterparts as they are composed of mainly light molecules and possess low spin-orbit interaction and hyperfine interaction [6]. PCPs are especially the better choice as their conjugation length is much higher than the small molecules and oligomers, leading to better transport properties and simpler fabrication process [7]. The ability to manipulate electron spin in organic molecules offers an alternative route to conventional spintronics, both from fundamental and technological points of view.

Magnetic field effects on OS were demonstrated already in the early '90s [8]. However, demonstration of successful spin injection and transport in lateral [9] and vertical [10] spin-valve devices and at the same time large OMAR and magneto-electroluminiscence (MEL) response in OS diode devices with non-magnetic electrode at room temperature under a small applied magnetic field (~10-100 mT) [11-13] caused resurgence of interest. Several experiments have so far successfully demonstrated efficient spin injection and transport in organic spin –valve structures and their successful operations, even at room temperature [14,15], paving the way for future commercial application. However, the reason for large OMAR effect in organic diodes have not been explained very successfully yet. Different models have been proposed to explain the observed effects [16-20] but differences between theory and experimental results still exist.

Our experiments with organic spin-valves [17] with RRP3HT as the spacers showed enhanced switching field than expected from $La_{0.7}Sr_{0.3}MnO_3$ and Co electrodes, indicating formation of magnetic domains in the structure. Also diodes with RRP3HT shows together with large OMAR there is a hysteretic effect of device resistance as a function of magnetic fields that indicates that OS might have some magnetic domain formation [21]. The magnetization of the OS needs to be thoroughly studied in order to fully understand different organic spintronic devices. In the present paper, we report the magnetic characterization of different OS systems together with their OMAR characteristics and correlate that OMAR in OS diodes is closely linked with the magnetic properties of respective OS.

In an earlier report [22] we chose the RRP3HT:PCBM blend as the model system to study the effect of electron-hole (e-h) recombination on the OMAR response of the device. And the results clearly showed that, with decreasing e-h pair formation probability, OMAR response decreases drastically. In our present experiment also, we have chosen regio-regular poly (3-hexyl thiophene) (RRP3HT), 1-(3-methoxycarbonyl)propyl-1-phenyl-[6,6]-methanofullerene (PCBM) and the RRP3HT:PCBM blend as the test systems to clarify whether there is any exact correlation between the magnetic and magneto-transport properties in the OS.

**Experimental**

The device structure used for the magneto-transport experiments is decribed earlier [22]. For the magnetic measurements, RRP3HT, either pure powder (as received) or drop-casted films from chloroform or dichlorobenzene (DCB) solution was used. The PCBM powder was measured as received. For the blend, 1:1 weight ratio of RRP3HT: PCBM was dissolved in DCB and films were made by drop-coating method on $SiO_2$ substrates to minimize any magnetic signal from the substrates. For all the films, a solution of density 5 mg/ml was used and the approximate thickness of the films were ~1 μm. The sample preparation was done in a nitrogen-filled glove-box and using anhydrous solutions. After fabrication, the films are transferred in a nitrogen atmosphere to the Superconducting quantum interface designed (SQUID) magnetometer. Temperature dependence of magnetization, M, was measured both after cooling the sample under zero field and then measuring the sample while heating under an applied field (ZFC) and during cooling the sample under the measuring field (FC) at temperatures between 5 – 300 K. Magnetic hysteresis curves were recorded in the field of B = ± 150 mT. All measurements in this case were made in dark. The external field B was always applied along the plane of the films.

For elemental analysis by the particle induced x-ray emission (PIXE) technique, about 15 mg of the polymer sample material was pressed to a pellet (diameter 13 mm). Pure graphite was used as backing material to minimize the amount of sample material [23]. The samples were irradiated with a 3 MeV proton beam from the Åbo Akademi MGC-20 cyclotron. The acquisition time was about 500 s with a beam current of 10 nA. All irradiations were performed in air to avoid heating and charge build-up. A strong ion luminescence from the irradiated spot on the polymer samples was observed in the beginning of the proton irradiation but faded away within a few seconds. This phenomenon indicates some changes in the molecular structure but does not affect the elemental concentration measurements. The radiation emitted from the sample during the irradiation was measured with an IGP X-ray detector for PIXE analysis. The integrated charge on the target needed for quantification was determined from measurements of light induced in air by the proton beam [24]. The obtained PIXE spectra were analyzed using the GUPIX software package [25]. The calibration was checked using the USGS granite CRM G-2. The procedure with the quality assurance has been described earlier [26].

**Results and Discussion**

Fig. 1 shows hysteretic behaviour of magnetoresistance for variation with magnetic field from 0 to +300 mT to -300 mT and back for a RRP3HT diode. The scan speed for the figure shown was 20 µT/sec. There is no evidence of such hysteretic behaviour in current-voltage characteristics and the device was stable for the whole measurement period [21]. This observation suggests presence of magnetic domains with long spin relaxation time in the diode. To study this phenomenon further we conducted magnetization measurement for the individual OS that show OMAR.

RRP3HT powder was wrapped in clean Teflon[©] [27] tape and were measured in the SQUID magnetometer in the temperature range of 5 – 300 K both in ZFC and FC directions with an applied field of 150 mT. The sample shows open magnetization vs. magnetic field (M-B) curves (Fig. 2) at all measured temperatures clearly indicating ferromagnetic (FM) ordering in this film even up to 300 K. Magnetization value increases sharply below 100 K. The ZFC and FC curves start to separate from each other below 200 K, indicating domain freezing behaviour. The magnetic hysteresis shows the saturation field to be ~100 mT. The measured saturation magnetic moment ($M_S$) value is 5.57 x $10^{20}$ $\mu_B$ $Kg^{-1}$ at 5 K and 3.98 x $10^{20}$ $\mu_B\_Kg^{-1}$ at 300 K. For films made with RRP3HT dissolved in chloroform and in DCB the $M_S$ value and the coercive field ($H_C$) changes drastically. For the film made with DCB, the $M_S$ value is 1.08 x $10^{20}$ $\mu_B$ /cc at 5 K and remains almost same till 300 K whereas for the film with chloroform $M_S$ value is much smaller i.e. 5.9 x $10^{18}$ $\mu_B$/cc at 5 K and it is almost impossible to measure the signal above 50 K. The $H_C$ value at 5 K also changes significantly from 16.5 mT for the Chloroform film to 4.5 mT for the DCB film. This result indicates that the film morphology plays a very important role in determining the FM ordering in PCPs. Fig. 2(a) shows the magnetic hysteresis loops of the RRP3HT films at room temperature together with the OMAR as a function of magnetic field at room temperature (Fig. 2(b)). The OMAR was measured using 100 µA current in a standard diode configuration with ITO and Al electrode and the measurement technique has been described previously in details [22].

Similar measurements were done on PCBM powder wrapped in clean Teflon[©] tape in the temperature range of 5 – 300 K in ZFC and FC direction. FM ordering was also observed in the whole temperature range. It is also observed that, FM ordering increases with decreasing temperature and the FC and ZFC curves starts to separate from each other below 50 K. The ZFC magnetization remains almost constant throughout the temperature range while the FC magnetization increases sharply below 50 K. The magnetic hysteresis shows (Fig. 2(c)) the saturation field to be ~100 mT. The measured saturation magnetic moment ($M_S$) value is 2.5 x $10^{20}$ $\mu_B$ $Kg^{-1}$ at 5 K and 1.99 x $10^{20}$ $\mu_B$ $Kg^{-1}$ at 300 K. Fig 2(d) shows the OMAR response of a PCBM diode device made similarly as the RR-P3HT diode device. Although the OMAR response is smaller in this case compared to the RRP3HT diode, clear OMAR signal was obtained from each device.

Next we studied the magnetization behaviour of the RRP3HT:PCBM blend and found that the 1:1 weight ratio of RRP3HT and PCBM is completely paramagnetic (PM) at all temperatures between 5 – 300 K. Fig. 2 (e) shows the M-B plot of a typical 1:1 blend of RRP3HT and PCBM at room temperature. The PM signal is also temperature independent and there is no difference of ZFC and FC signal. Upon closer inspection, it is observed that FM in the blend still exists but it is suppressed by the large PM signal. Earlier also, PM response of the same blend was reported using light induced electron spin resonance spectroscopy and the result was explained as separation of charge carriers into different paths on the polymer and fullerene blends. However, this is the first time when we observe this large PM signal in the blend in the dark. This present result suggests that a ground state charge transfer complex is formed. Figure 2(f) shows the OMAR response of a diode made from the blend, orders of magnitude smaller than that of a RRP3HT or PCBM diode.

In a RRP3HT:PCBM blend, PCBM is embedded in the conjugated polymer and acts as a strong electron acceptor whose lowest unoccupied molecular orbital (LUMO) lies below the excitonic state. It has been shown earlier that upon irradiation, electron is transferred from a polymer chain to a fullerene molecule within femtosecond time scale [28], whereas the electron back transfer is much slower and this elecron transfer results in a charge-separated state. Recently, Krinichnyi [29] showed upon irradiation, this charge separated state can give rise to huge PM signal and the effective PM susceptibility of the generated polarons and fullerene anion-radicals, is inversely proportional to the probability of their recombination. However, we observe strong PM signal from the 1: 1 blend of RRP3HT and PCBM already in the dark, implying that a ground state charge transfer complex is formed in the RRP3HT:PCBM blend leading to free carriers. The PM signal arising from these free carriers suppress the individual FM ordering in RRP3HT and PCBM and we observe only a large PM signal.

In order to understand the FM ordering in the polymers RRP3HT and PCBM, we performed PIXE analysis of the samples to search for magnetic impurities. Figure 3 shows a typical PIXE spectrum, it has clear peaks from S, Ni and Br. Sulfur can be found in the RRP3HT backbone and bromine in the ends of the polymer chain, the other elements are considered to be impurities. Elemental concentrations and statistical uncertainties for the samples are shown in Table 1.

After examining the PIXE data, it was observed that both in RRP3HT and PCBM samples the amount of magnetic impurities like Fe, Ni, Co, Mn etc. are very small. Traces of Ni were found in both the polymers while Fe was the main magnetic impurity in PCBM. Earlier [30-32] very weak FM was observed in $C_{60}$ molecule and the FM ordering was attributed to the presence of Fe impurities. In RRP3HT the magnetization value is ~5.5 x $10^{20}$ $\mu_B$/kg while the Ni impurity amount is ~12 □g/g (Table 1). In PCBM, the magnetization value is ~4.5 x $10^{20}$ $\mu_B$/kg while the Fe impurity amount is ~15 □g/g (Table 1). Now in different compounds and complexes Ni spin magnetic moment varies from 2.9 – 4.0 $\mu_B$/atom while Fe spin moment varies from 5.1 – 5.7 $\mu_B$/atom. So, the magnetic moment arising from Ni impurity (in RRP3HT) can be ~3.5-4.5 x $10^{20}$ $\mu_B$/kg while that from Fe impurity (in PCBM) can be ~9 x $10^{20}$ $\mu_B$/kg. Both the value matches quite well with the experimentally obtained results indicating that the magnetic moment can actually arise from the magnetic impurities present in the samples. However, the spatial separation between the dilute magnetic impurities in the polymer matrix is too large to have any direct interaction between them. Hence, we believe that the FM interaction is mediated via an intermediary that gives rise to long-range FM ordering, as previously shown in EuO where the $O_2$ ligands mediates the super exchange interaction [33]. Comparison between fullerene (C60) and PCBM shows that presence of hydrogen atoms, i.e hyperfine interaction, in PCBM increases FM ordering (results not shown here). Also, OMAR is not observed in C60 diodes [34]. This indicates that the intermediary is most likely the hyperfine interaction for long-range FM ordering in OS. However, detailed nuclear magnetic resonance (NMR) study of hydrogen atoms is needed to further elaborate the role of hyperfine interaction in determining FM ordering in OS.

Another important point that should be taken into account here is the strong morphology dependence of the RRP3HT saturation moment. Although the amount of magntic impurity is same in both the films the saturation moment varies by almost 2 orders of magnitude depending on the film morphology. So, it becomes evident that depending on domain size FM interaction can vary significantly. This leads to the conclusion that trapped carrier mediated FM intraction, as is usually observed in metal-oxide systems [35], could also play an important role here. In all the OS we see a very small spontaneous magnetization value (almost zero). As we turn-on the magnetic field, the FM impurities in OS becomes polarized and their spins are oriented along the external field direction (Kondo effect). The spins of the trapped carriers in the immediate vicinity of the impurity atom also get polarized and form a magnetic cluster. Overlapping of such clusters can give rise to long range FM ordering.

As shown in Fig. 2, we have observed that when a sizable FM ordering in a material exists, we always observe a large OMAR response in the diodes with the respective material. However, upon destroying the FM ordering, we observe a large PM signal and at the same time OMAR response disappear. This observation clearly challenges our present understanding of the OMAR effect and suggests that OMAR might have a correlation with the FM ordering in OS, observed in presence of small magnetic impurities and sizeable concentration of hydrogen atoms. Details of the magneto-transport mechanism can be better understood after studying the electric-field induced magnetization study, currently underway.

In conclusion, we have observed FM ordering in a pi-conjugated polymer RRP3HT and PCBM small molecules but 1:1 blend of these two materials show large PM signal. PIXE analysis of the individual materials suggests the presence of magnetic impurity which can account for the observed magnetic moment. However, this dilute magnetic impurity is unable to produce long range FM ordering and is most likely mediated by the hyperfine field arising from the large number of hydrogen atoms in the hydrocarbons. Successful separation of charge carriers in a RRP3HT:PCBM blend destroys the FM ordering and gives rise to a PM signal. This implies that charge transfer complex is formed in the blend already in the ground state. Corresponding magneto-transport measurement reveals that OMAR effect can only be observed in the FM phase of the material and it vanishes in the RRP3HT:PCBM blend suggesting that FM ordering is responsible for large OMAR effects.

**Acknowledgement:**

The authors gratefully acknowledge the Wihuri Foundation and financial support from the Academy of Finland projects 116995 and 107684 through the Centre of Excellence Programme. Planar International Ltd. is acknowledged for the patterned ITO substrates.

**Figure captions:**

Fig. 1. Magnetoresistance as a function of magnetic field (B) at 300K for RRP3HT diode measured with 1μA current showing resistance hysteresis behaviour. B was scanned from 0 to +300 mT to -300 mT and back at a scan speed of 20 μT/s.

Fig. 2. Magnetization as a fuction of magnetic field (B) at 300 K for (a) RRP3HT powder, (c) PCBM powder and (e) RRP3HT:PCBM blend film on $SiO_2$ substrate. % Magnetoresistance as a function of magnetic field at 300 K for diode devices with (b) RRP3HT, (d) PCBM and (f) RRP3HT:PCBM solar cells with ITO as the hole and Al as the electron injector.

Fig. 3. A typical energy spectrum of regio-regular poly(3-hexyl thiophene) (RRP3HT) showing presence of Ni among different impurities.

Figures:

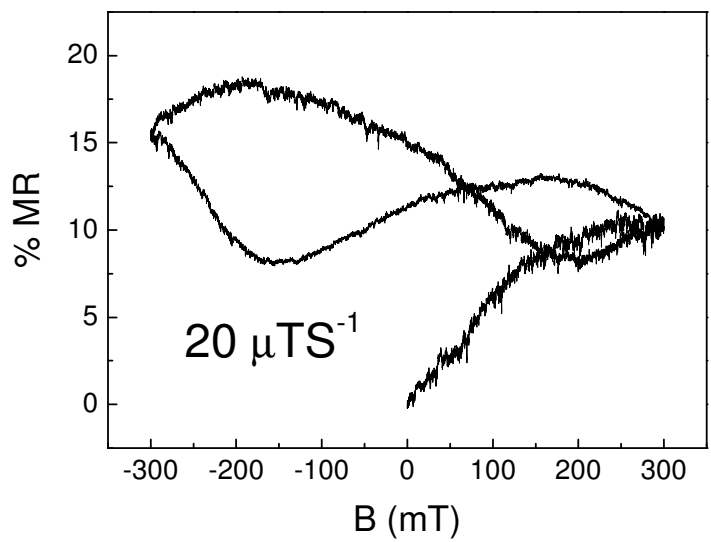

Figure 1. S. Majumdar et al.

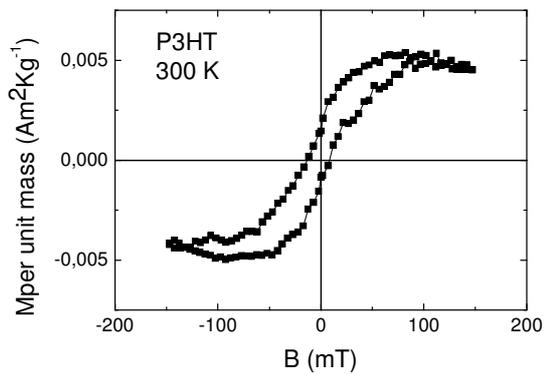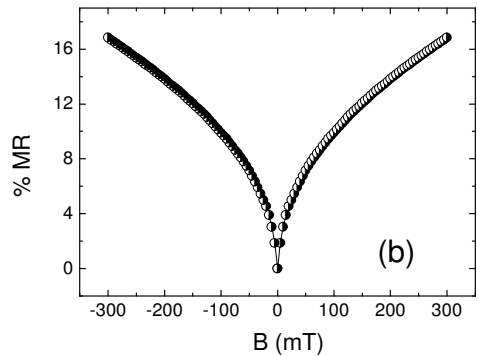
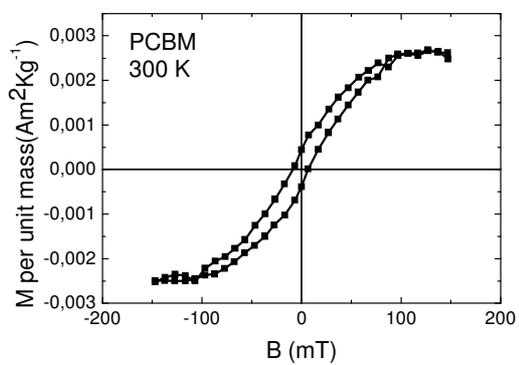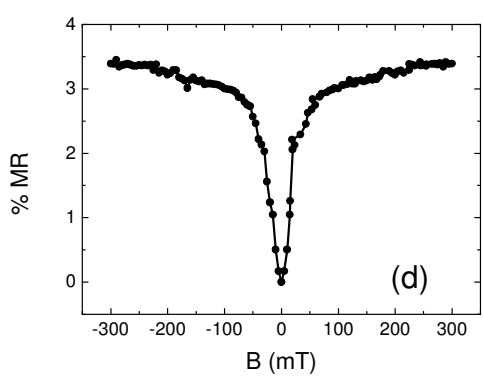
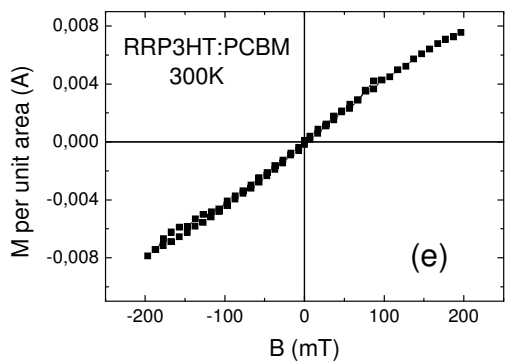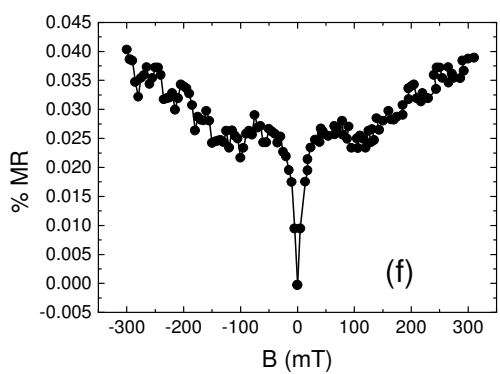

Figure 2. S. Majumdar et al.

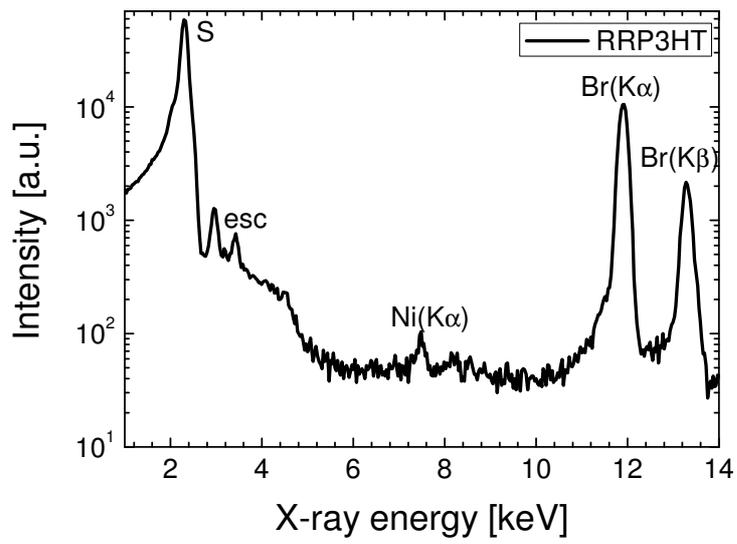

Figure 3. S. Majumdar et al.

Table – 1. Elemental concentrations in the materials given in µg/g of dry weight. Typical statistical errors in % and limit of detection (LOD) in µg/g are in the columns to the right.

| Elemental content ↓ | RRP3HT | PCBM | Error (%) | LOD |
|---|---|---|---|---|
| S | 180360 | bdl | 0.2 | 190 |
| Br | 3975 | bdl | 0.4 | 1.4 |
| Mn | bdl | bdl | - | 8.7 |
| Fe | bdl | 15 | 20 | 4.9 |
| Ni | 12 | 1.7 | 14 | 1.2 |

*bdl – below detection limit*